\documentclass[aps]{revtex4}
\usepackage{graphicx}

\begin{document}

\hfill Applied Physics Report 2003--25

\title{van der Waals Interactions of Parallel and 
Concentric Nanotubes}

\author{Elsebeth Schr\"{o}der}
\author{Per Hyldgaard}\thanks{E-mail: \texttt{hyldgaar@fy.chalmers.se}}

\affiliation{Department of Applied Physics, Chalmers University of Technology and
G\"{o}teborg University, SE--412 96 Gothenburg, Sweden}

\date{June 11, 2003}

\begin{abstract}

For sparse materials like graphitic systems and carbon 
nanotubes the standard density functional theory (DFT) faces 
significant problems because it cannot accurately describe 
the van der Waals interactions that are essential to the
carbon-nanostructure materials behavior. While standard implementations
of DFT can describe the strong chemical binding within an isolated,
single-walled carbon nanotube, a new and extended DFT implementation
is needed to describe the binding between nanotubes. We here provide 
the first steps to such an extension for parallel and concentric nanotubes
through an electron-density based description of the materials coupling 
to the electrodynamical field. We thus find a consistent description 
of the (fully screened) van der Waals interactions that bind the 
nanotubes across the low-electron-density voids between the nanotubes,
in bundles and as multiwalled tubes. 
\end{abstract}
%Nanotubes; van der Waals Interactions; Dispersion 
%Interactions; Density Functional Theory Calculations

\maketitle

\section{Introduction}

Soft and sparse materials constitute an important challenge
for our first-principle quantum-physical account of structure
and dynamics. These structures are defined by voids of
very-low-electron density regions across which material
binding is communicated exclusively via the coupling to 
the electrodynamic fields, the so-called dispersion or
van der Waals (vdW) forces. Traditional implementations of
density functional
theory (DFT) cannot account for this type of interaction 
because the density functionals (DF) are defined as local
or semilocal and fail to describe the truly nonlocal
nature of the electrodynamic coupling. Nevertheless,
the traditional approaches are imperative for accurate predictions of the
binding within the dense regions of finite electron
densities. The many opportunities in biological systems,
in organic-molecular liquids and in the carbon 
nanostructures motivate a continuous search for a 
consistent combination of traditional DFT and vdW corrections.
This biophysics and nanotechnology perspective
motivated our recent proposal for a vdW-DF for 
layered systems~\cite{Tractable,GraphiteJapan,GraphiteSubmit,condmat}.

The interaction in nanotube bundles, ropes~\cite{yu}, and within 
multiwalled nanotubes~\cite{cumings}, as addressed here, 
represents another
type of sparse-matter problem with a nanotechnological 
relevance~\cite{dai}.  Nanotube bundles are often directly produced 
and/or extracted to align the nanotube for Raman spectroscopy 
whereas multiwalled nanotubes are produced in many fabrication
techniques. Both the nanotube bundles and the concentric
multiwalled nanotubes are stabilized by a competition between 
the kinetic-energy repulsion and the long-ranged van der Waals 
binding. At the relevant binding distances
(3 to 3.5 {\AA}  or 6 to 7 Bohr radii) the traditional
first-principle DFT calculations cannot account for the 
combined interaction. The issues and challenges are similar
to and closely related to the corresponding problem of 
ab-initio calculations of the binding in graphite~\cite{condmat}. 

An accurate quantum-physics account of the binding in nanotube
bundles and in multiwalled nanotubes must include a calculation
of the dispersion forces defined by the electrodynamics coupling.
To illustrate the importance of these van der Waals contributions
we here use a simple model description of the single-tube
electrodynamics response fitted to DFT calculation of the
nanotube electron system and response. While simple, our approach 
improves the traditional Hamaker-type of estimates 
\cite{garifalco} in several ways: (i) it offers a method for
a first-principle calculation of the interaction strength,
(ii) it includes the important local-field screening
effects as directly defined by the anisotropic spatial
variation in the nanotube electron density (which we here 
obtain from DFT); and (iii) it consistently accounts for
the net anisotropic dielectric response that describes
extended molecules. In this manner we establish a link 
between the microscopic description (based on accurate DFT calculations
of the electron density variations and static susceptibility
\cite{condmat}) and the phenomenologic descriptions, and we
provide a method to test the limit of applicability. In a wider sense 
our nanotube study also provides an indirect insight on how the 
electron-density variation determines the optical response 
of large extended (organic) molecules as used, for example,
in liquid-crystal displays~\cite{LDres}.

In the present analysis of the nanotube van der Waals interaction 
we focus on interaction strength at intermediate separations when 
the interaction is defined by the dipolar contribution 
\cite{condmat,Unified}. This emphasis permits a number of 
very important simplifications. With a set of 
approximations we succeed in using the high degree of
symmetry which exists in the case of parallel and concentric
nanotubes to split the interaction into an effective coupling
constant (given by one frequency integration) and a nanotube
geometry factor. In the worst case the computation is thus reduced
to just a two-dimensional spatial integration, as shown in the
following.
Moreover, both for the cases of concentric nanotubes and 
of identical parallel nanotubes we further succeed 
in expressing this general integral fully or partially in 
terms of spatial function for additional efficiency gain.

\section{The nanotube electrodynamic response}

%%%%%%%%%%%%%%%%%%%%%%%%%%%%%%%%%%%%%%%%%%%%%%%%
\begin{figure}
\begin{center}
\scalebox{0.35}{\includegraphics%[width=70mm]
{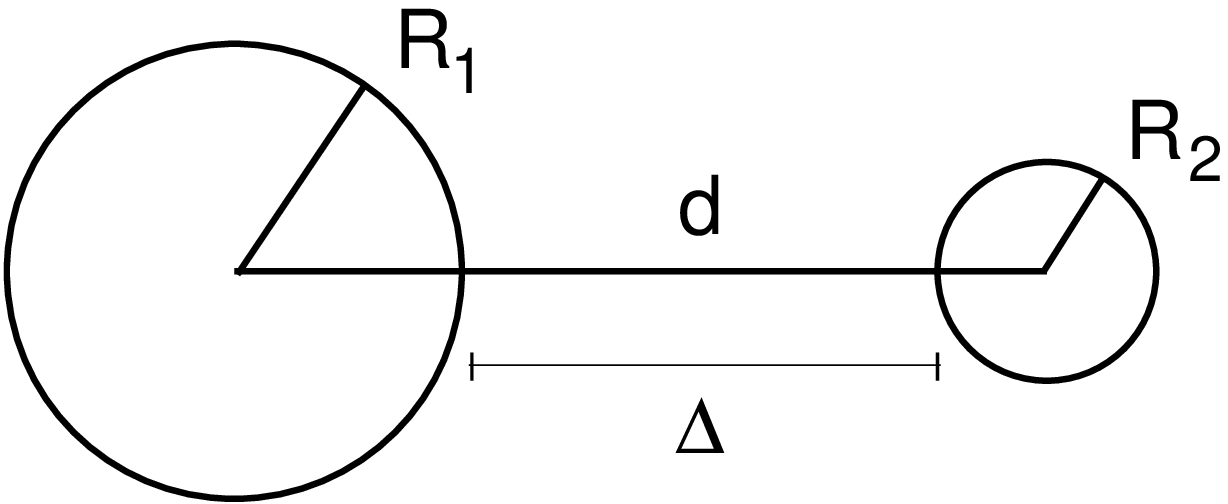}}\\[1cm]
\scalebox{0.35}{\includegraphics%[width=70mm]
{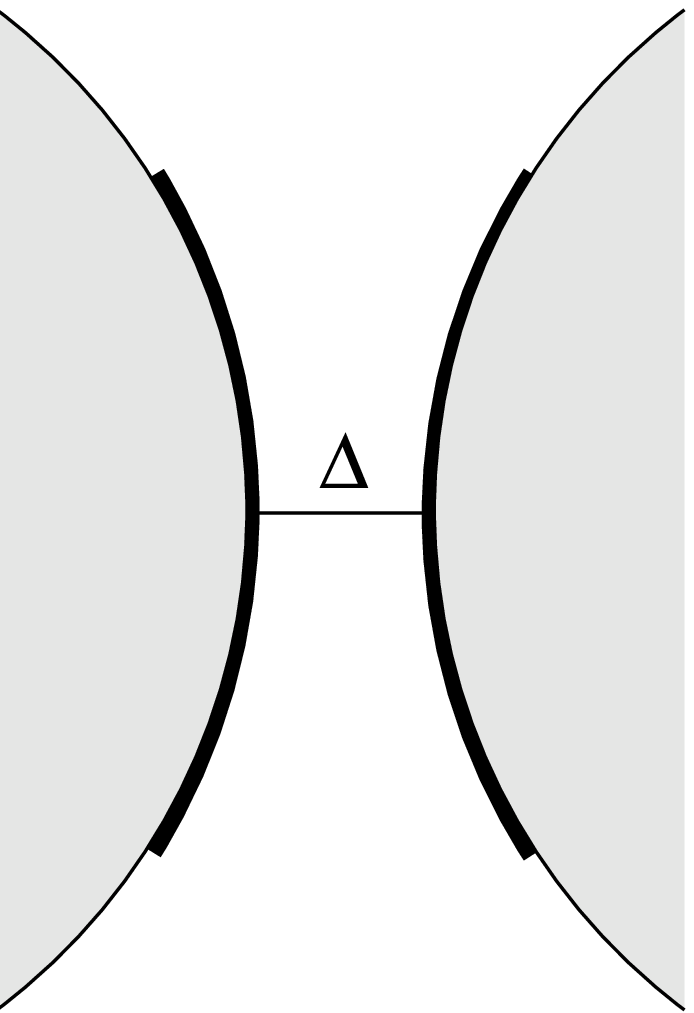}}
\caption{\label{figsketch} Sketch of parallel nanotubes. 
{\it Top panel:\/} Tubes with radii $R_1$ and $R_2$, seen along the 
$z_1$ and $z_2$ axes. {\it Bottom panel:\/}
A closer look at the closest interaction region. Whereas all parts of
a nanotube in principle interact with all parts of the other nanotube
the strongly attractive character of the dipole-dipole interaction 
causes most of the contribution to the collective interaction to come 
from within the
region near the line of closest contact.
}
\end{center}
\end{figure}
%%%%%%%%%%%%%%%%%%%%%%%%%%%%%%%%%%%%%%%%%%%%%%%%

Our calculations of the
physical intertube attraction are based on accurate first-principle 
DFT calculations of the nanotube electron density, the
overall, nanotube electrostatic response, and an approximate treatment 
of the nanotube electrodynamical response, local-field effects,
and resulting intertube attraction. The approach is 
described in Ref.~\cite{schhyl03} and will be summarized here.

Adapting the plasmon-pole model of Ref.~\cite{Unified},
we approximate the local microscopic electron response (for
a sub-atomic-scale element of the nanotube wall) by the bare 
dynamic susceptibility 
$
\chi_0\left(n(\mathbf{r}),u,u_0\right)=n(\mathbf{r})/(u^2+u_0^2)
$
at (complex) frequency $u$. 
Hartree atomic units are used with symbols $a_0$ for the Bohr radius and Ha 
for the hartree ($\approx 27.21$ eV).
The nanotube electron density $n(\mathbf{r})$ (at position
$\mathbf{r}$) is determined directly from 
first-principle DFT calculations, and the 
value of the effective frequency cut-off $u_0$ is 
obtained by comparing first-principle calculations
of the static nanotube susceptibility to a 
calculation including the local-field effects produced by  
the model of $\chi_0$.

{}From this model of $\chi_0$ we obtain a 
corresponding effective susceptibility tensor 
$\chi_{\mbox{\scriptsize eff}}$
which describes the ratio of the locally induced polarization 
to the externally applied electric field.
We describe the response of a nanotube to 
an applied electric field using a local
cylindrical coordinate system $(s,\theta,z)$
with the $z$-axis along the nanotube axis
and $\mathbf{s}$ a vector perpendicular to the $z$-axis.
Approximating the nanotube electron density $n(\mathbf{r})$ by its 
radially averaged value $n(s)$
the local effective susceptibility is given by the relation 
\begin{equation}
\chi_{\mbox{\scriptsize eff}}\left[n(s)\right](u)
\mathbf{E}_{\mbox{\scriptsize applied}}
=
- \chi_0\left(n(s),u,u_0\right)
\nabla \phi(\mathbf{s},u)
\end{equation}
where the local electric field $-\nabla \phi(\mathbf{s},u)$ is 
given by charge conservation
$\nabla \cdot\left\{ \left(  1+4\pi\chi_0\right)
\nabla \phi  \right\}=0$.
In general, the effective susceptibility tensor
$\chi_{\mbox{\scriptsize eff}}$ will be anisotropic with off-diagonal elements.
However, when transforming the tensor $\chi_{\mbox{\scriptsize eff}}$
into cylindrical coordinates all off-diagonal elements vanish
and only the diagonal elements  
$\hat{\chi}_{\mbox{\scriptsize eff}}^{\beta}$, $\beta=s,\theta,z$
are left.

%%%%%%%%%%%%%%%%%%%%%%%%%%%%%%%%%%%%%%%%%
\begin{figure}
\begin{center}
\includegraphics[width=80mm]
{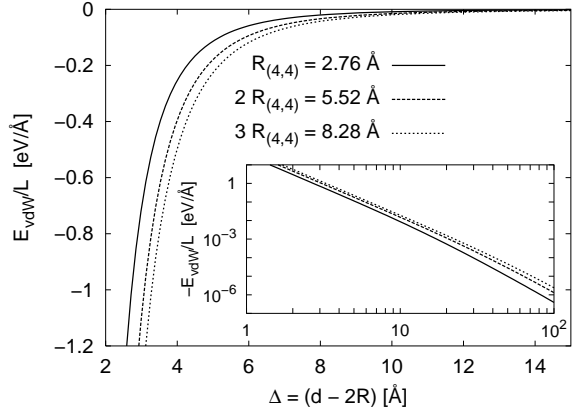}
\caption{\label{figevdwdeltalin} 
Van der Waals interaction of pairs of parallel nanotubes
with pairwise identical radii, as a function of 
smallest separation $\Delta=d-2R$. The attraction of
large nanotubes (radius $3R_{(4,4)}$) is greater than that of 
small tubes (radius $R_{(4,4)}$) at the same $\Delta$. 
Large nanotubes have less curvature and thus at a given
distance $\Delta$ a pair of large tubes has a larger amount of charge
at distance approximately $\Delta$, yielding a larger total binding
per tube length. 
}
\end{center}
\end{figure}
%%%%%%%%%%%%%%%%%%%%%%%%%%%%%%%%%%%%%%%%%

For an external electric field of strength $E_0(u)$, oriented 
perpendicular to the nanotubes or along the nanotubes,
we may factorize the electric potential as 
$\phi(\mathbf{s},u,u_0)=-E_0(u)W(s,u,u_0)\cos\theta$ 
and solve $\nabla \cdot\left\{ \left(  1+4\pi\chi_0\right)
\nabla \phi  \right\}=0$.
The components of the effective susceptibility then become~\cite{schhyl03}
$\hat{\chi}_{\mbox{\scriptsize eff}}^{s}=
\chi_0 \partial_s W$,
$\hat{\chi}_{\mbox{\scriptsize eff}}^{\theta}=
\chi_0 W/s$, and 
$\hat{\chi}_{\mbox{\scriptsize eff}}^{z}=
\chi_0 $.

To proceed with an efficient determination of the
internanotube van der Waals interactions we exploit
the special form of the nanotube electron density.
In the thin-wall approximation 
we approximate the 
radial behavior of the effective susceptibility 
$\hat{\chi}_{\mbox{\scriptsize eff}}^{\gamma}(s,u,u_0) \to
\tilde{\chi}^{\gamma}(R,u,u_0)\delta(s-R)$,
by a weighted radial delta function such that
$
\int d^2 s \,\hat{\chi}_{\mbox{\scriptsize eff}}^{\gamma}(s,u,u_0) \equiv
2\pi R \,\tilde{\chi}^{\gamma}(R,u,u_0)$.

These approximations are the basis for derivation of the
nanotube-nanotube interaction.

\section{The nanotube-nanotube van der Waals interaction}

For a pair of parallel nanotubes at center-to-center separation 
$d$ (Fig.~\ref{figsketch}) we define two sets of local cylindrical
coordinate systems with origos separated by $d$, and with indices 1 and 2
referring to the two nanotubes and their local coordinate systems.
Within the dipole-dipole approximation we then find the
van der Waals energy
\begin{eqnarray}
E_{\mbox{\scriptsize vdW}}&=&-\frac{1}{2\pi}
\int_0^\infty du\,  \mbox{Trace}
\left\{\hat{\chi}_1T_{12}(d) \hat{\chi}_2 T_{21}(d)\right\}
\label{Evdw}\\
&=&
-\frac{L}{2\pi}
\sum_{\beta,\gamma=s,\theta,z}
\left( 
\int_0^\infty du\, 
\tilde{\chi}^{\beta}(R_1,u,u_0^{(1)})
\tilde{\chi}^{\gamma}(R_2,u,u_0^{(2)})R_1R_2
\right)
\nonumber \\
&& \times
\int_0^{2\pi} d\theta_{1}\int_0^{2\pi} d\theta_{2}I_{\beta\gamma}
(\theta_{1},\theta_{2},R_1,R_2)
\label{Evdw2}
\end{eqnarray}
with the geometry terms
\begin{eqnarray}
I_{\beta\gamma}(\theta_{1},\theta_{2},s_1,s_2)
&=&\frac{1}{L}
\int_{-L/2}^{L/2} dz_1 \int_{-\infty}^{\infty}dz_{21}
T_{12}^{\beta\gamma}(d)T_{21}^{\gamma\beta}(d)
\nonumber\\[0.6em]
&=&
\int_{-\infty}^{\infty}dz_{21}\left(T_{12}^{\beta\gamma}(d)\right)^2
\end{eqnarray}
and the dipole coupling $T_{12}=
-\nabla_1\nabla_2\left|\mathbf{r}_2-\mathbf{r}_1\right|^{-1}$
between elements of the two nanotubes.  
We made use of the thin-wall approximations 
 to carry out the $s_1$ and $s_2$ integrations.
The geometry terms $I_{\beta\gamma}$ are of a form for
which the $z_{21}$ integration can be easily be carried 
out~\cite[(3.241.4)]{gradsh}:
\begin{equation}
\int_{-\infty}^{\infty}dz_{21} \frac{z_{21}^{\mu-1}}{(s_{21}^2+z_{21}^2)^5}
= \frac{\Gamma\left(\frac{\mu}{2}\right)\Gamma\left(5-\frac{\mu}{2}\right)}{24}
s_{21}^{-(10-\mu)}
\end{equation}
with $\mu=1$, 3, or 5, and $s_{21}=|\mathbf{s}_2-\mathbf{s}_1|$.

The cut-off frequencies $u_0^{(1)}$ and 
$u_0^{(2)}$ for the two nanotubes depend in principle on the 
nanotube radius, but as found in Ref.~\cite{schhyl03} the macroscopic 
susceptibility is not very sensitive to the
value $u_0$. In practice we thus take the large-nanotube value
of $u_0$ for all nanotubes, $u_0^{(1)}=u_0^{(2)}=0.30$ Ha. 

%%%%%%%%%%%%%%%%%%%%%%%%%%%%%%%%%%%%%%%%%%%
\begin{figure}
\begin{center}
\includegraphics[width=80mm]
{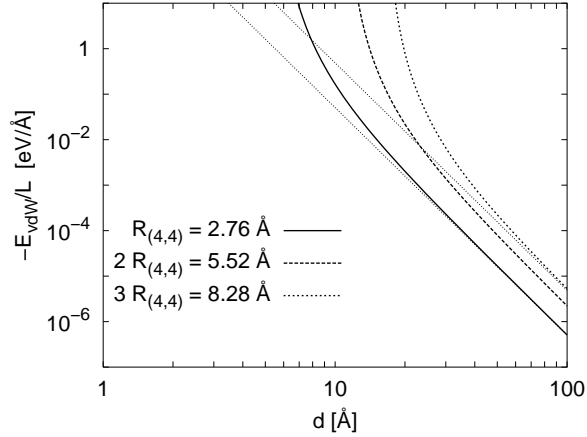}
\caption{\label{figevdwdlog} 
Van der Waals interaction of pairs of parallel nanotubes
as a function of center-of-tube to center-of-tube distance $d$.
Gray lines are the asymptotes 
$E_{\mbox{\scriptsize vdW}}/L\sim-C_5/d^5$ 
with 
$C_5=9\pi^2
J_{\mbox{\scriptsize tot}}(R,R,u_0,u_0)/64$
for $R=R_{(4,4)}$ and $R=3R_{(4,4)}$.
}
\end{center}
\end{figure}
%%%%%%%%%%%%%%%%%%%%%%%%%%%%%%%%%%%%%%%%%%%

For parallel nanotubes we can thus reduce the original simultaneous
six-dimensional spatial ($\mathbf{r}_1$, $\mathbf{r}_2$) and 
one-dimensional ($u$) frequency integral to a sum of
terms each factorized into a one-dimensional frequency integral
times a two-dimensional ($\theta_1$, $\theta_2$) spatial 
integral. The frequency integrals, now decoupled from
the spatial integrals, do not depend on the nanotube
separation $d$ and are therefore identical to the frequency factors
defined in Ref.~\cite{schhyl03}
\begin{equation}
J_{\beta\gamma}(R_1,R_2,u_0^{(1)},u_0^{(2)})=\int_0^\infty du 
\tilde{\chi}^{\beta}(R_1,u,u_0^{(1)})
\tilde{\chi}^{\gamma}(R_2,u,u_0^{(2)})R_1 R_2\,\,.
\end{equation}
As shown earlier~\cite{schhyl03}, the special case of concentric 
nanotubes ($d=0$) further reduces the number of spatial dimensions to 
be integrated.

In general it is not possible to carry out all of the 
nine spatial integrals $I_{\beta\gamma}(\theta_{1},\theta_{2},s_1,s_2)$
to express them in terms of known functions.
The complications, however, are reduced for certain special cases
such as concentric nanotubes~\cite{schhyl03} and, as shown below,
for parallel nanotubes of equal radius.

To emphasize the effect of the cylindrical symmetry
on the van der Waals interaction (\ref{Evdw2}) we would like to 
pursue analytic calculations as far as possible.
For this reason we shall assume that 
the effective radial and tangential susceptibility 
are identical, $\tilde{\chi}^{s}\equiv \tilde{\chi}^{\theta}$.
The error of this approximation is largest for small 
frequencies $u$, and is at most 10\%.
Thus, with 
$J_{ss}=J_{\theta s}=J_{s\theta}=J_{\theta\theta}$, 
$J_{sz}=J_{\theta z}$, and $J_{zs}=J_{z\theta}$,
the van der Waals energy per nanotube length becomes
\begin{eqnarray}
\frac{E_{\mbox{\scriptsize vdW}}}{L}
&=&
-\frac{1}{2\pi}\int_0^{2\pi} d \theta_{1}\int_0^{2\pi} d \theta_{2}
\big\{
J_{ss}\left(I_{ss}+I_{s\theta}+I_{\theta s}+I_{\theta\theta}\right)
\nonumber \\
&&\mbox{}+ J_{sz}\left(I_{sz}+I_{\theta z}\right)
+ J_{zs}\left(I_{zs}+I_{z\theta}\right)
+J_{zz}I_{zz}
\big\}\,\,.
\label{eqeji}
\end{eqnarray}
In Eq.~(\ref{eqeji}) each sum of $I$-terms has the same functional form,
\begin{equation}
I_{\mbox{\scriptsize tot}}(\theta_{1},\theta_2,R_1,R_2)
=\frac{9\pi}{128}\left((R_2\cos\theta_2+d-R_1\cos\theta_1)^2+
(R_2\sin\theta_2-R_1\sin\theta_1)^2\right)^{-5/2}
\end{equation}
times an integer factor:
$I_{ss}+I_{s\theta}+I_{\theta s}+I_{\theta\theta} = 3I_{\mbox{\scriptsize tot}}$, 
$I_{sz}+I_{\theta z}= 5I_{\mbox{\scriptsize tot}}$, 
$I_{zs}+I_{z\theta}= 5I_{\mbox{\scriptsize tot}}$, 
$I_{zz}= 19I_{\mbox{\scriptsize tot}}$.
This is an important observation, as it enables
the factorization into frequency and spatial integrals
not only for each term in the sum of~(\ref{eqeji}),
but indeed for the full sum of~(\ref{eqeji}). 
If we therefore define an effective frequency integral 
$J_{\mbox{\scriptsize tot}}
=3J_{zz}+5J_{zs}+5J_{sz}+19J_{ss}$ the van der Waals energy per length
can be written as the product
\begin{equation}
\frac{E_{\mbox{\scriptsize vdW}}}{L}
=-\frac{1}{2\pi}
J_{\mbox{\scriptsize tot}}(R_1,R_2,u_0^{(1)},u_0^{(2)})
\int_0^{2\pi} d \theta_{1}\int_0^{2\pi} d \theta_{2}
I_{\mbox{\scriptsize tot}}(\theta_{1},\theta_2,R_1,R_2)
\label{eqejisimple}
\end{equation}

%%%%%%%%%%%%%%%%%%%%%%%%%%%%%%%%%%%%%%

\begin{figure}
\begin{center}
\includegraphics[width=80mm]
{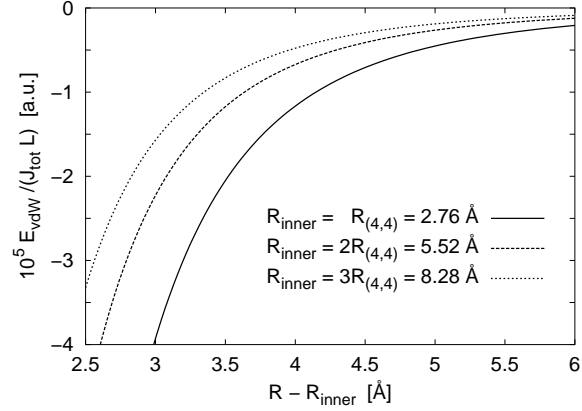}
\caption{\label{figconc} 
The geometry part of the van der Waals interaction of concentric nanotubes
for fixed inner-tube radius $R_{\mbox{\scriptsize inner}}$ and as a 
function of outer-tube radius $R$.
}
\end{center}
\end{figure}

%%%%%%%%%%%%%%%%%%%%%%%%%%%%%%%%%%%%%%

Eq.~(\ref{eqejisimple}) is our result for 
arbitrary-size parallel nanotubes, including concentric nanotubes. 
The two- and one-dimensional 
spatial and frequency integrals
involved in (\ref{eqejisimple}) are already a considerable 
simplification compared to the original $6+1$ dimensional
integral, and can be numerically solved. 

For identical nanotubes, $R_1=R_2=R$, the
two spatial integrals in (\ref{eqejisimple}) further reduce
 to a single integral over the variable
$\xi =(\theta_2-\theta_1)/2$
\begin{eqnarray}
\frac{E_{\mbox{\scriptsize vdW}}}{L}
&=&-\frac{9\pi}{64}
J_{\mbox{\scriptsize tot}}(R,R,u_0,u_0)
\times \nonumber \\
&&\times
\int_{-\pi/2}^{\pi/2} d\xi
\,\,{}_2F_1\left(\frac{5}{4},\frac{7}{4};1;
\left(\frac{4dR\sin\xi}{d^2+4R^2\sin^2\xi}\right)^2\right)
(d^2+4R^2\sin^2\xi)^{-5/2}
\end{eqnarray}
where ${}_2F_1$ is a hypergeometric function.
Replacing one integral in this way with a well-known, tabulated
special function speeds up the numerical evaluation
of $E_{\mbox{\scriptsize vdW}}$ and enhances the accuracy.

In Figure~\ref{figevdwdeltalin} we plot $E_{\mbox{\scriptsize vdW}}$
as a function of the smallest distance between elements of the
nanotubes, $\Delta=d-R_1-R_2=d-2R$ for three different radii.
In this and the following figures we have assumed a constant value 
of $u_0=0.30$ Ha.

In the asymptotic limit, $d>> R_i$, 
\begin{eqnarray}
\frac{E_{\mbox{\scriptsize vdW}}}{L}&=&
-\frac{9\pi^2}{64}
\frac{J_{\mbox{\scriptsize tot}}(R_1,R_2,u_0^{(1)},u_0^{(2)})}{d^5}
\times \nonumber \\[0.6em]
&&\times\left\{1+\frac{25}{2}\frac{(R_1^2+R_2^2)/2}{d^2}
+\frac{3675}{32}\frac{(R_1^4+4R_1^2R_2^2+R_2^4)/6}{d^4}
+{\cal O}\left(\frac{1}{d^6}\right)\right\}
\label{expd}
\end{eqnarray}
we recover to lowest order
the traditional London-theory macroscopic $d^{-5}$ behavior~\cite{mahanti} 
of two thin, parallel cylinders, albeit with different higher-order correction
terms because nanotubes are hollow, not solid, cylinders.
The limit of the expansion~(\ref{expd}) is illustrated by the plots in 
Figure~\ref{figevdwdlog} for the identical
tubes presented in Figure~\ref{figevdwdeltalin}. 

Figure~\ref{figconc} shows our results for another important 
special case, namely the concentric nanotube
system~\cite{schhyl03}. When the two local nanotube coordinate systems
coincide ($d=0$) both of the two spatial integrals in  
(\ref{eqejisimple}) can be solved 
exactly~\cite[(2.584.58)]{gradsh} to give
$E_{\mbox{\scriptsize vdW}}$ in terms of
the first and second complete Legendre elliptic
integrals $K(k)$ and $E(k)$
\begin{eqnarray}
\frac{E_{\mbox{\scriptsize vdW}}}{L}
&=&
-\frac{3\pi}{32}\frac{ J_{\mbox{\scriptsize tot}}(R_1,R_2,u_0^{(1)},u_0^{(2)})}%
{(R_2+R_1)^{3}(R_2-R_1)^{2}}
\times
\nonumber \\
&&\times
\left\{4\frac{R_2^2+R_1^2}{(R_2-R_1)^2}
E\left(\frac{2\sqrt{R_1R_2}}{R_1+R_2}\right)
-K\left(\frac{2\sqrt{R_1R_2}}{R_1+R_2}\right)\right\}.
\label{eqconc}
\end{eqnarray} 
In Figure~\ref{figconc} the geometry part of (\ref{eqconc}),
$E_{\mbox{\scriptsize vdW}}/( J_{\mbox{\scriptsize tot}} L)$, 
is shown.

We stress that the repulsive nanotube-nanotube
interaction, resulting from the kinetic-energy repulsion, 
is not treated within our formalism.

\section{Conclusions}

In summary, we report a quantum-physics calculation of the 
vdW binding in nanotube bundles and in concentric (multiwalled)
nanotubes which is based directly on first-principle calculations 
of the electron density and on the electron density response.
While present DFT implementations cannot account for the intertube
interactions we have thus identified a method for a consistent
combination with ab-initio calculations. Together with our
recent progress for the layered systems our results indicate
that a combined vdW-DF is feasible and hence promises 
an integration for a full quantum-physics account of
the sparse and soft materials used for example in biophysics
and nanotechnology.

\section{Acknowledgments}

This work was supported by the
Trygger Foundation, the Swedish Research Council (VR), and the 
Swedish Foundation for Strategic Research (SSF).


\begin{thebibliography}{99}

\bibitem{Tractable}
H.~Rydberg, B.~I.~Lundqvist, D.~C.~Langreth and M.~Dion, {\it Phys.~Rev.~B\/} 
\textbf{62}, 6997 (2000).

\bibitem{GraphiteJapan} B.~I.~Lundqvist et al., 
{\it Surface Sci.\/}  \textbf{493}, 253 (2001).

\bibitem{GraphiteSubmit} 
H.~Rydberg, N.~Jacobson, P.~Hyldgaard, S.~I.~Simak, B.~I.~Lundqvist,
and D.~C.~Langreth, {\it Surface Sci.\/} 
\textbf{532--535}, 606 (2003).

\bibitem{condmat} H.~Rydberg, M.~Dion, N.~Jacobson, E.~Schr{\"o}der, 
P.~Hyldgaard, S.~I.~Simak, D.~C.~Langreth, and B.~I.~Lundqvist,
{\it Van der Waals Density Functional for Layered Structures\/},
\texttt{cond-mat/0306033}~.

\bibitem{yu} M.-F.~Yu, B.~S.~Files, S.~Arepalli, and R.~S.~Ruoff,
{\it Phys.~Rev.~Lett.\/} {\bf 84}, 5552 (2000).

\bibitem{cumings} J.~Cumings and A.~Zettl, {\it Science}  
\textbf{289}, 602 (2000).

\bibitem{dai} H.~Dai, {\it Surf. Sci.\/} {\bf 500},  218  (2001).

\bibitem{garifalco} L.~A.~Girifalco, M.~Hodak, and R.~S.~Lee, 
{\it Phys.~Rev.~B\/} \textbf{62}, 13104 (2000).

\bibitem{LDres} E.~Schr\"oder, {\it Phys.~Rev.~E\/} \textbf{62}, 8830 (2000).

\bibitem{Unified} E.~Hult, H.~Rydberg, B.~I.~Lundqvist,
and D.~C.~Langreth, {\it Phys.~Rev.~B\/} {\bf 59}, 4708 (1999).

\bibitem{schhyl03} E.~Schr\"oder and P.~Hyldgaard, {\it Surface Sci.\/} 
\textbf{532--535}, 880 (2003).

\bibitem{gradsh} I.~S.~Gradshteyn and I.~M.~Ryzhik, \textit{Table of Integrals,
Series, and Products}, 5th edition (Academic Press, San Diego, 1979).

\bibitem{mahanti} J.~Mahanty and B.~W.~Ninham, {\it Dispersion Forces\/}
(Academic Press, London, 1976).

\end{thebibliography}
\end{document}